# Exploring the Correlation between Solvent Diffusion and Creep Resistance of Mg-Ga HCP Alloys from High Throughput Liquid-Solid Diffusion Couple

Jingya Wang[a,b,d], Guanglong Xu[c], Xiaoqin Zeng[a], Javier Llorca[b,d], Yuwen Cui[c]*

[a] National Engineering Research Center of Light Alloy Net Forming and State Key Laboratory of Metal Matrix Composite, Shanghai Jiao Tong University, 200240 Shanghai, PR China

[b] IMDEA Materials Institute, Getafe Madrid, Spain

[c] Tech Institute for Advanced Materials & College of Materials Science and Engineering, Nanjing Tech University, Nanjing, 211816, China

[d] Departament of Materials Science, Polytechnic University of Madrid/Universidad Politécnica de Madrid, E. T. S. de Ingenieros de Caminos, 28040 - Madrid, Spain.

* Corresponding author, ycui@njtech.edu.cn

## Abstract

The liquid-solid diffusion couple technique, supported by phenomenological analysis and nano-indentation tests, is proposed on account of the relatively low melting points of Mg to explore the diffusion mobility and creep deformation. The potential of this strategy is demonstrated in Mg-Ga hcp alloys where Ga solute (i.e. impurity) and Mg solvent diffusions in hcp Mg-Ga alloys were both unveiled. It was followed by mapping the compressive creep behavior via nanoindentation along the composition arrays within the same Mg-Ga couple sample. The compressive creep resistance of Mg-Ga hcp alloys increased with the Ga content, and this enhancement was similar to the one found in Mg-Zn alloys and superior to the one reported in Mg-Al alloys though Al is a slower impurity diffuser in hcp-Mg than Zn and Ga. Thereby, the solvent diffusion and its variation with the composition, rather than the solute diffusion, was suggested to govern the creep properties at high temperatures and low stresses.

**Keywords:**

Liquid-solid diffusion couple, hcp Mg-Ga alloys, diffusion mobility, creep resistance, Mg solvent diffusion

# 1. Introduction

Magnesium is the lightest structural material that has a lot of potential applications in engineering as a result of the good castability, high specific-stiffness, and excellent electrical conductivity [1,2]. However, Mg-based materials have comparatively poor strength and inferior creep resistance at elevated temperature. Thus, diffusion, creep and stress-rupture data are very relevant to design Mg alloys for high-temperature applications. Nevertheless, magnesium and its alloys are highly reactive, present low melting points and have inferior weldability, which make diffusion experiments very laborious. The creep test of magnesium alloys is not straightforward either, being time-consuming and costly [3].

The diffusion properties greatly influence a wide range of phenomena in Mg alloys such as solidification and oxidation kinetics, phase equilibria, precipitation and high-temperature deformation [4,5]. Technical difficulties in diffusion experiments on magnesium alloys mostly lie in casting dilute Mg alloys and assembling solid-state diffusion couple due to poor weldability. As a result, reliable diffusion data of magnesium alloys are limited. Recently, Zhong and Zhao [6–8] took advantage of quick liquid phase formation at elevated temperatures and proposed a non-trivial liquid-solid diffusion couple (LSDC) technique. They succeeded in extracting efficiently the impurity diffusion coefficients of a number of elements in hcp-Mg.

The conventional creep test is rather time consuming, costly and thus limited too. Recently, nanoindentation creep tests have been used for assessing the rate-controlling and time-dependent plastic response of a material under stress. In particular, nanoindentation creep tests have been successfully employed to study rate-dependent process at the microscale in Mg and Mg alloys [9–11]. Moreover, dynamical nanoindentation creep tests have become an advanced high throughput screening approach once it is combined with the diffusion couple and multiple experimentation [12].

The diffusion properties of materials are strongly linked with diffusion creep mechanisms at low stresses and elevated temperatures (Nabarro-Herring (N-H) creep [13,14] and Coble creep [15]). For other processes, creep is operated via distinctive mechanisms like dislocation glide, cross slip and grain boundary sliding [14].

For a long time, it has been very frequently observed that the activation of creep energy is similar to the activation energy for lattice self-diffusion for a large number of materials [16,17]. Particularly, the creep rate of the N–H creep and the diffusion-assisted dislocation climb creep has been linked to lattice diffusion due a fact that the high temperature steady-state creep of crystalline solids possesses an constant-valued activation energy roughly equivalent to that of self-diffusion [16,17]. Recently, it was suggested that high diffusion of solutes [7] in hcp-Mg was responsible for the low creep resistance of Mg alloys. Nevertheless, Ca is a fast impurity diffuser in hcp-Mg, and this mechanism cannot explain the enhancement in creep resistance of Mg-Ca alloys [7]. The activation energy for impurity diffusion quite often differs by orders of magnitude from the self-diffusion activation energy. For magnesium alloys, Al, Sn, Y, Mn and Gd impurities diffuse slower in hcp-Mg than the self-diffusion of hcp-Mg, while Ca, Ga, Ce and Zn impurities diffuse faster [6,18]. There are no specific data on the solvent diffusion in Mg binary alloys. It therefore calls a need to evaluate or measure the solvent diffusion and raises a question of which type of diffusion is better suited to determine the creep resistance of Mg-rich alloys.

Nowadays, increasing attention has been paid to the Ga-containing Mg alloys because of their electrochemical activity, particularly as sacrificial anodes in seawater batteries [19] and of their improved corrosion behavior in medical implants [20]. Accordingly, the objectives of this work are to put forward high throughput strategy based on the liquid-solid diffusion couple technique, supported by phenomenological analysis and nanoindentation tests, to extract the diffusion mobility and compressive creep resistance. The strategy is demonstrated on Mg-Ga alloys and further applied to shed light on if the solvent diffusion could be served as the indicator of creep properties.

## 2. Experimental procedure

2.1 Fabrication of liquid-solid diffusion couple

Mg-1.6at.%Ga and Mg-15at.%Ga alloys were prepared from pure magnesium (99.95 wt.%) and Mg-Ga master alloys by induction melting under an argon atmosphere.

The pure magnesium cylinders and blocks of binary alloys cut from the ingots were solid-solution treated under vacuum in quartz capsules at 673 K for 5 days followed by water quenching.

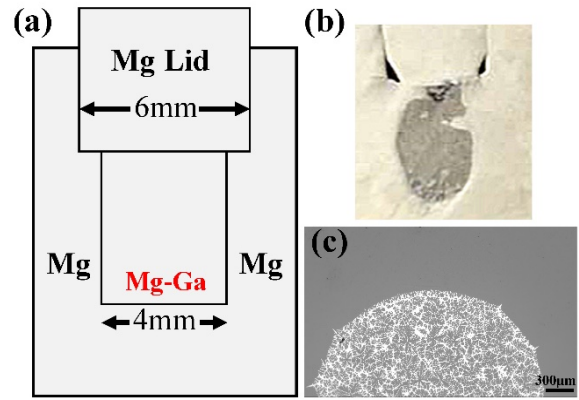

Fig. 1(a) Schematic of liquid-solid diffusion couple; (b) Optical micrograph of Mg-Ga LSDC; (c) Representative SEM image of the Mg/Mg-15at.%Ga LSDC after annealing at 848K for 24 hours.

The pure magnesium can was prepared by drilling a 4 mm diameter hole with a depth of 15 mm at the center. The L/S diffusion couples were manufactured by inserting the Mg-15 at.% Ga cylinder into the hole, and were further tightened and sealed by compressing from the top with a pure Mg lid of 6 mm diameter. The LSDC diffusion couple assembly process was performed in an argon glove box, as illustrated schematically in Fig. 1(a). The severe plastic deformation and heat resulted from the compression led to binding the lid with the Mg can tightly and prevented air from contaminating the S/L diffusion couples when they were placed out of the glovebox. After long-term high temperature interdiffusion annealing between 748 and 848 K, the couples were quenched into water to keep the high-temperature microstructures, and no sign of obvious leakage or oxidation was found. This was further confirmed by subsequent compositional analysis and microstructural characterization. The diffusion couples were then sectioned, mounted, and polished by standard metallographic techniques. The microstructure of diffusion zone was observed by scanning electron microscopy and the local composition was analyzed by electron microprobe analysis

(EPMA, JEOL 8900).

The consistency and reliability of the diffusion behavior of the Mg-Ga alloys from the S/L diffusion couples were afterwards validated by comparison to standard solid/solid (S/S) diffusion couples prepared at 773 K. The S/S diffusion couples, constrained within a composition range up to 1.6 at.% Ga, were assembled by joining two disks in a bonding holder by tightening the screws. All diffusion couples made in this work are listed in Table 1.

Table 1 Heat treatment and preparation method for the diffusion couples

| Couple No. | Temperature (K) | Time (hours) | Preparation method |
| --- | --- | --- | --- |
| 1 | 748 | 120 | Solid/Liquid |
| 2 | 773 | 120 | Solid/Liquid |
| 3 | 773 | 72 | Solid/Solid |
| 4 | 798 | 72 | Solid/Liquid |
| 5 | 823 | 72 | Solid/Liquid |
| 6 | 848 | 24 | Solid/Liquid |

2.2 Nanoindentation tests

Nanoindentation creep and quasi-static indentation tests were carried out using the commercial TriboIndenter TI950 nanoindentation system from Hysitron, Inc. (Minneapolis, MN), equipped with a diamond Berkovich tip. Constant load indentation creep tests were used to characterize the creep behavior whilst quasi-static indentation tests were conducted to measure the hardness and the modulus of Mg alloys as a function of Ga content. The nanoindentation creep tests were performed with a loading rate of 50 μN/s, until a prescribed load of 12 mN was attained at room temperature. The maximum depth was dependent on the material response (including its creep response). Afterwards, the creep behavior was examined during a creep dwell period of 900s under nanoDMA conditions with a frequency of 200 Hz and a displacement amplitude of ~

2.3 nm. After the holding period, the load was released to zero at an unloading rate of 500 μN/s. The indentation creep constitutive parameters can be extracted from the above curves following the protocol detailed [21,22]. The nanoindentation tests were performed under load control with a loading and unloading time of 5 s and a 2 s holding time at the maximum load of 12 mN. Hardness and modulus values were computed from the unloading portion of the load-displacement curves by applying the Oliver-Pharr method [23]. Additional nanoindentation tests were performed in Mg/MgAl diffusion couple following a procedure in Ref. [24] to extract the composition-dependent hardness and modulus in the Mg-Al alloys for comparison to those in the Mg-Ga alloys.

**3. Diffusion and mobility of Mg-Ga alloys**

Figure 2(a) shows the composition profiles of the Mg-Ga L/S diffusion couples annealed at 748, 773, 798 and 848 K, respectively. The composition profiles on the left side were measured from the hcp-Mg solid solution phase within the diffusion region, whilst those on the right correspond to the liquid phase diffusion. The solubilities of Ga in Mg at 748, 773, 798 and 848 K were 2.47, 2.39, 1.69 and 1.23 at.% Ga, respectively, in good agreement with the Mg-Ga binary phase diagram [25]. The composition profiles from the S/S diffusion couple annealed at 773 K is plotted in Fig. 2(b).

Extracting the diffusion coefficients directly from the dual-phase diffusion couple is not straightforward. A novel direct optimization scheme developed by Zheng et al [26] was utilized in this work to extract the diffusion mobility for the hcp Mg-Ga binary phase directly from the experimental profiles of L/S diffusion couples. The scheme was coded using the MATLAB interface toolbox [26] to communicate with the Thermo-Calc and DICTRA softwares. The thermodynamic parameters for Mg-Ga system employed in this work were taken from the widely accepted description assessed by Meng et al. [27], that could precisely represent the experimental phase diagram and thermodynamic properties. There are no available diffusion mobility parameters for liquid phase of Mg-Ga alloys and the diffusion mobility of liquid Mg-Al alloys [28] was used instead, assuming that Ga has the similar properties to Al from the viewpoint of diffusion and that the diffusion mobility for the liquid phase has negligible influence on the optimization process. The diffusion mobility parameters corresponding to the self-diffusion in hcp-Mg determined by Bryan et al. [29] were used in this work because they enable to describe the majority of diffusional experimental data. Because no stable hcp-Ga exists, the self-diffusion of hypothetical hcp-Ga was estimated on the basis of the semi-empirical self-diffusion relations [30,31], i.e.:

$$Q_B = RT_m(K + 1.5V), \quad \text{(Eq 1a)}$$

$$D_0 = 1.04 \times 10^{-3} Q_B a^2, \quad \text{(Eq 1b)}$$

where $T_m$ stands for the melting temperature, $K$ is a constant that depends on the crystal structure and is assumed to be 5.5 for hcp metals, $V$ represents the valence and $a$ is the lattice parameter. The theoretical melting temperature of hcp-Ga was roughly deduced on the basis of Miedema's empirical model [32] from the intrinsic atomic properties computed using first-principles calculation by Wang. et al. [33], while the lattice parameters were directly taken from the calculation by Wang. et al. [33]. The end-member for the impurity diffusion of Mg in hcp-Ga was assumed to equivalent to the self-diffusion of hcp-Ga.

The mobility parameters for the impurity diffusion of Ga in hcp-Mg were optimized by fitting to the experimental data and they are listed in Table 2. Note that no binary interaction parameter was yielded since the solubility of Ga in hcp-Mg is very limited. The experimental composition profiles from the S/L diffusion couples at 748, 773, 798, and 848 K, as well as the S/S diffusion couple at 773 K, are compared with the simulated values in Fig. 2(a) and Fig. 2(b), respectively. It is noted that the simulated composition profiles are in good agreement with the experimental data. The interdiffusion coefficients of hcp Mg-Ga alloys are plotted in Fig. 2(c) as a function of Ga content, and they increase with the Ga content. Our calculations not only well describe the composition and temperature dependence of the interdiffusion coefficients, but are also in good agreement with the interdiffusion coefficients extracted from the S/S diffusion couple by Sauer-Freise method [34].

Table 2 The optimized diffusion mobility parameters in this work

| Mobility | Parameter (J/mol) | Reference |
| --- | --- | --- |
| $\phi_{Mg}^{Mg}$ | -125077-88.17*T | [29] |
| $\phi_{Mg}^{Ga}$ | -90123-109.09*T | This work |
| $\phi_{Ga}^{Ga}$ | -90123-109.09*T | This work |
| $\phi_{Ga}^{Mg}$ | -126885-76.86*T | This work |

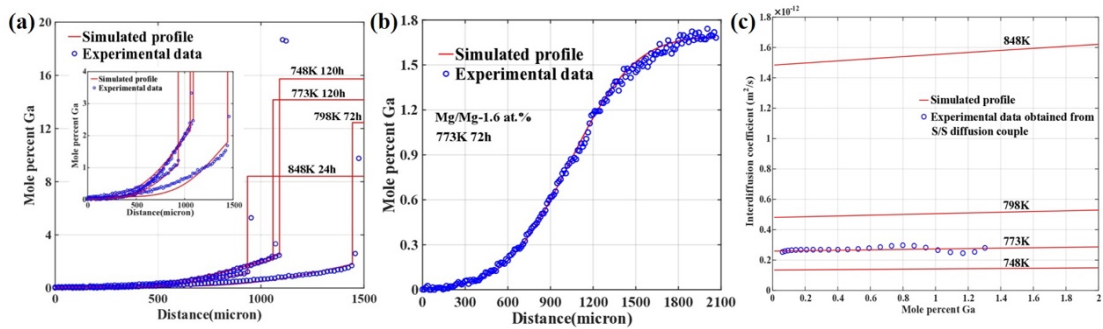

Fig. 2 (a) Experimental and simulated composition profiles of LSDC at four temperatures; (b) Experimental and simulated composition profiles of solid-solid diffusion couple; (c) Interdiffusion coefficients as a function of the Al composition for the hcp phase of the Mg-Ga system obtained from this study in comparison with the results extracted from the solid/solid diffusion couple.

The impurity diffusion coefficient of Ga in hcp-Mg is plotted in Fig. 3(a), compared with the values extracted from the S/S diffusion couple in this study using the Hall method [35], and those from Staloukal et al. [36] using the residual activity method, as well as from the first-principles calculation by Zhou et al. [37]. These data can be reasonably expressed by an Arrhenius relation with a frequency factor $D_0=9.67*10^{-5}$ m$^2$/s and a diffusion activation energy Q=126.89 kJ/mol, as shown by the red line in Fig. 3(a). The predictions obtained from the optimized mobility parameters nicely reproduce the experimental data extracted in the present work, but are significantly higher than the experimental values from Staloukal et al. [36]. Nevertheless, it was considered that these latter experimental data underestimate the mobility parameter due to the low rate of γ-rays emission that led to poor statistics in the radioactive assay of the concentration [38]. The first-principle calculations of the Ga impurity diffusion coefficients overestimate slightly the experimental results. Similar trends were observed in the case of Gd, Ce, Ca, Al and Sn [7,8,39]. The impurity diffusion of Ga in hcp-Mg is further compared in Fig. 3(b) with various alloying elements (inc. Al, Ca, Ga, Sn, Y, Mn and Zn) [6–8,40] with a reference to the Mg self-diffusion. Apparently, Ga is much alike Zn and diffuses faster compared to the self-diffusion of Mg, while Al is among those slower impurity diffusers like Sn and Y.

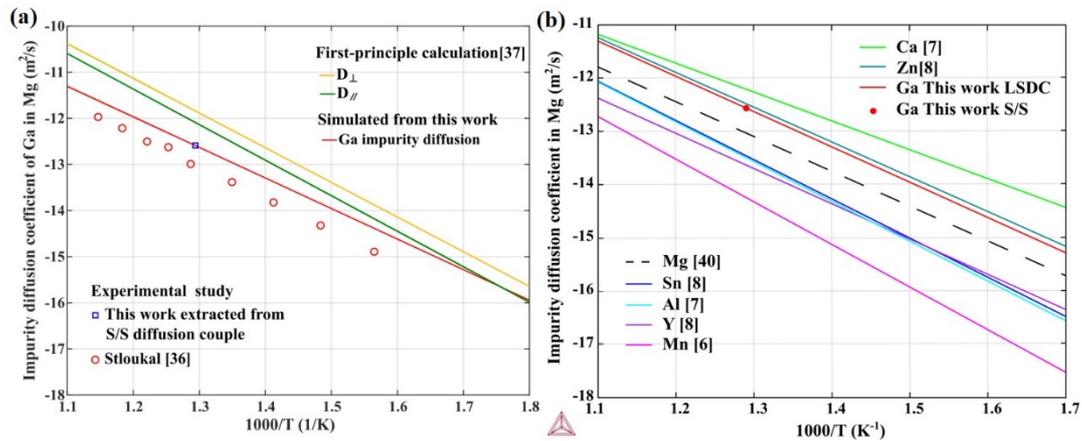

Fig. 3 (a) Impurity diffusion coefficients of Ga in Mg. The experimental results [36] are shown as symbols and the first-principles calculations [37] are shown as solid lines for comparison. (b) Impurity diffusion coefficients of solute X (Al, Ca, Ga, Sn, Y, Mn and Zn) [6–8,40] are compared with reference to the Mg self-diffusion coefficient.

## 4. Micro-mechanical properties of Mg-Ga alloys

4.1 Nanoindentation hardness

The effect of Ga content on the indentation hardness and Young's modulus is shown in Fig. 4. Both increase with the Ga content, although the former exhibits a much stronger composition-dependence. The monotonic increase of the hardness with the solid solute content in Mg alloys has been investigated in the case of Mg-Al [41], Mg-Zn [42] and Mg-Y [43] alloys, although the hardness was measured with the Vickers tester. In this work, the nano-indentation hardness ($h_{ind}$) of Mg-Ga alloys was screened for a large range of compositions in the LSDC samples, and the solid solution strengthening due to Ga can be expressed as

$$h_{ind} = 0.159 c_{Ga} + 0.837 \ (GPa), \quad (Eq\ 2)$$

where $c_{Ga}$ is Ga content in at.%, and the strengthening factor was determined to be 0.159 (GPa/at.%), as shown in Fig. 4. The large effect of Ga on the solid solution strengthening of hcp Mg can be preliminarily ascribed to the large difference in the atomic radius between Ga and Mg, which lead to large lattice distortions that interact strongly with dislocation slip.

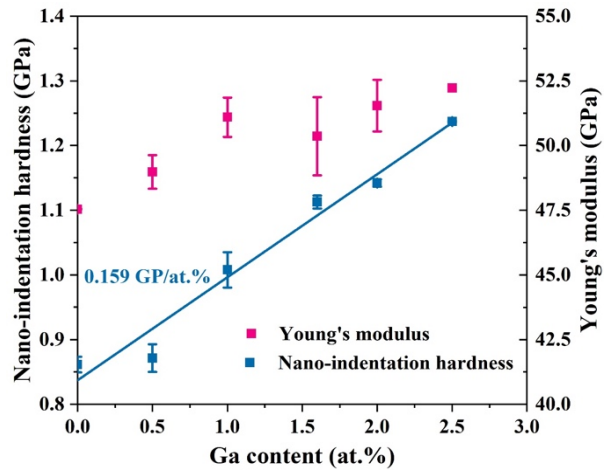

Fig. 4 Effect of Ga content on the hardness and elastic modulus of the Mg-Ga alloys.

4.2. Compressive creep resistance

The creep deformation at room temperature was investigated from the indentation creep curves for various Mg-Ga alloys, see the microstructure and indentations along the Ga composition gradient in Fig. 5(a). The creep depth in the indentation creep curves in Fig. 5(b) increased rapidly at the beginning of the test and it was followed by a slow creep dwell period where the secondary stage -characterized by an notable apparent steady-state rate- was finally reached [12,44]. It is apparent that the creep depth increases as the Ga content decreases as illustrated in Fig. 5(b) as a result of solid solution hardening. Moreover, the steady-state strain rate generally decreased when the

Ga content increased, and hence infers that the addition of Ga improves the compressive creep resistance. Close inspection, however, reveals that the enhancement of in the compressive creep resistance by alloying with Ga, i.e. the efficacy of the Ga strengthening, is slowed down as the content goes beyond 1.0 at% Ga.

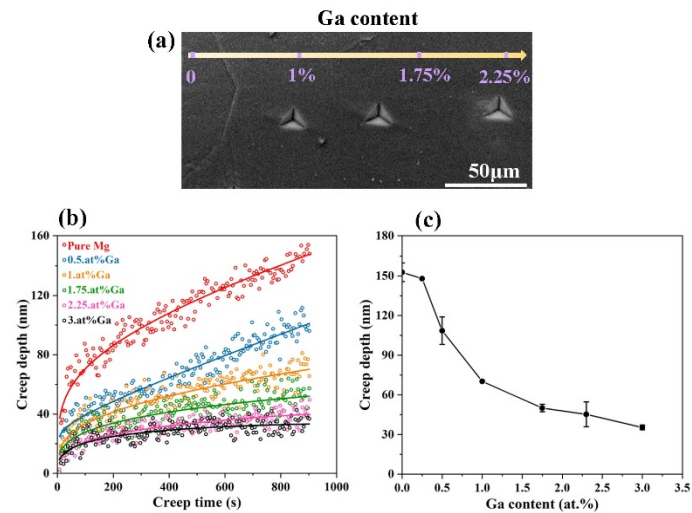

Fig.5 (a) Microstructure and indentations along the Ga composition gradient; (b) Creep curves of Mg-Ga alloys with various Ga contents; (c) Creep depth as a function of Ga content.

The strain rate sensitivity (SRS), *m,* and the activation volume extracted from the creep curves are plotted as a function of Ga content in Fig. 6 to provide an insight into the creep mechanisms in Mg-Ga alloys. They indicate that the addition of Ga reduces the SRS of Mg alloys from 0.054 for pure Mg to 0.012 for Mg-3at.%Ga alloys, while the activation volume increases from $38.4b^3$ for pure Mg to $81.6b^3$ for Mg-3at.%Ga alloys. In this study, the indentation creep tests were performed in the interior grain with the indent depth around ~ 700 nm at the maximum load, and grain boundaries had a negligible influence on the creep process. Generally, diffusional creep dominates the creep process at high temperatures ($\geq T_m$) under very low stress [4]. Therefore, it is presumed that the dislocation slip and/or twining should control the creep behavior at room temperature under high stress, which are the conditions of our experiments. The increase of the compressive creep resistance is mainly contributed to the pinning effect of solid solute Ga atoms on the dislocation motion during the rate-dependent deformation. It was reported that the creep resistance is improved by the addition of Al, Ca and Y [9,45,46] and Zn [44] due to the drag effect of solid solution and precipitation on the dislocations. The strain rate sensitivity for pure hcp-Mg was determined to be around 0.054 in the present study with an activation volume of 38.4$b^3$, thereby suggesting that dislocation cross-slip or dislocation slip were likely to be the dominant processes in this case [47]. A similar SRS < 0.1 for Mg was obtained by Somekawa et.al [11] using indentation creep tests, where dislocation slip was supposed to dominate. With the increase of the Ga, the SRS decreased while the activation volume increased.

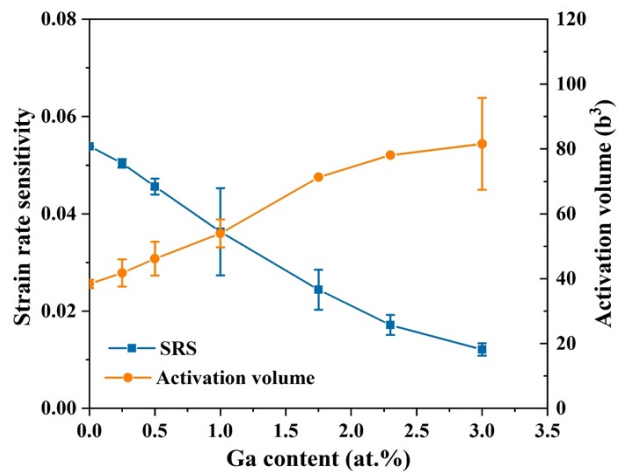

Fig.6 Strain rate sensitivity (m) and activation volume for creep deformation of Mg-Ga at room temperature as a function of Ga content.

## 5. Correlation of solvent diffusion and compressive creep resistance of Mg-Ga alloys

The activation energy for creep provides a useful basis to correlate the diffusion of metals and alloys with their creep behavior. Dislocation climb is the rate controlling mechanism for creep at low stresses and elevated temperatures whereas non-dislocation-based diffusional creep is generally dominant at high temperatures (T≈ 0.75$T_m$ or above) and very low stresses. The activation energies of these two types of creep are considered to be close to that for lattice diffusion, so is its compositional dependence.

The solvent atoms tend to be affected and interacted with the solutes as the solute content increases in dilute alloys [48]. Depending on the nature of interaction with the solute, the lattice diffusion of solvent could be accelerated or slowed down. For dilute Mg-Ga alloys, the solvent diffusivity of Mg can be written as a function of the solute content as

$$D_{Mg}^{*Mg-Ga} = D_{Mg}^{*}(1 + b_0 X_{Ga} + \cdots), \tag{Eq. 3}$$

where $b_0$ is a linear enhancement factor that is related to the various jump rates of the solvent atoms in the neighborhood of the solute [49,50]. From a phenomenological point of view, $b_0$ is equivalent to the regular interaction parameter, described by the Redlich-Kister polynomial in the spirit of the CALPHAD approach. The graphic representation of the diffusion coefficients of the Mg solvent in the Mg-X (Zn, Ga and Al) binaries is depicted by the highlighted portion in Fig. 7 (a). Note that the interaction parameters were omitted during the calculation for a better view of the relationships of various diffusion coefficients of solvent, solute and through the process of interdiffusion.

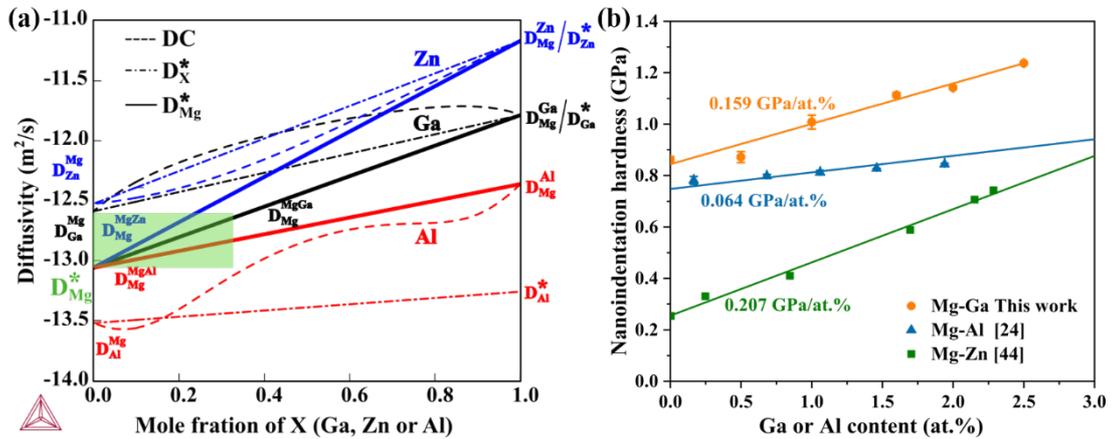

Fig. 7 (a) Graphic representation of the diffusion coefficients of Mg-X (Al, Ga and Zn) binaries calculated at 773 K; (b) Nano-indentation hardness of the Mg-X (Al [24], Ga and Zn [44]) solid solution alloys.

As learned from Fig. 7(a), the solvent diffusion of Mg is enhanced by all three solutes in the three binaries, and increases from Al to Ga and then to Zn with linear enhancement factors $b_0(Al)=1.7132$, $b_0(Ga)=3.3096$, and $b_0(Zn)=5.2414$, respectively. In essence, this contrasts with the fact of the solute diffusion where Zn and Ga diffuse faster than the self-diffusion of Mg while Al acts as a diffusion-slowing solute in hcp-Mg instead. It turns out that the solvent diffusion of Mg diffusion, rather than the impurity diffusion of solute, could better explain the experimental observation that the compressive creep resistance and nano-indentation hardness were all improved noticeably by alloying hcp-Mg with Al, Ga and Zn [44], as shown in Fig. 7(b). Consistently, our hardness tests also show that the enhancement efficacy of solid solution strengthening is strong with Zn and Ga, i.e. Zn and Ga provide stronger strengthening than Al though it is slower impurity diffuser in hcp-Mg. On the contrary, Au addition to Pb profoundly increases the solvent diffusion of Pb, which surprisingly deteriorates the creep resistance [51]. This is likely because Au is an interstitial diffusor in Pb.

## 6. Summary

The diffusion behavior in hcp Mg-Ga alloys was investigated by the liquid/solid diffusion and solid/solid diffusion couples. The diffusion mobility of hcp Mg-Ga binary alloys has been extracted by a direct optimization technique through fitting to the experimental composition profiles of the liquid/solid diffusion couple. The mobility parameters provide an appropriate representation of various diffusion coefficients of solute, solvent and interesting diffusion couple experiments.

Furthermore, the mechanical properties of Mg-Ga alloys were investigated by means of nanoindentation tests in the liquid/solid diffusion couple. The addition of Ga increases the hardness and improves the compressive creep resistance of Mg alloys at room temperature, evidencing a significant solid solution strengthening by alloying Mg with Ga. This mechanism is also responsible for the reduction in the strain rate sensitivity with the Ga content.

The compressive creep resistance and nano-hardness of Mg-Ga hcp alloys increases with Ga content by a similar amount to that reported in Mg-Zn and by a higher amount to that reported in Mg-Al alloys though Al is a slower impurity diffuser in HCP-Mg than Zn and Ga. The linear enhancement factors of three solutes in hcp-Mg are $b_0(Zn)=5.2414$, $b_0(Ga)=3.3096$, and $b_0(Al)=1.7132$, respectively. The solvent diffusion, rather than the impurity diffusion, is therefore suggested to correlate to the creep properties at high temperature and low stress.

It should be finally noted that the combined technique of liquid/solid diffusion couple supported by phenomenological analysis and nano-indentation tests could be widely applicable to investigate the mechanical properties and deformation mechanisms for a wide range of compositions, phase and microstructures using high-throughput strategies.

**Acknowledgements**

This work was supported by the Natural Science Funds of China [Grant No. 51571113 and 51825101]. G.X. was funded by the Natural Science Fund of China [Grant No. 51701094]. JLL acknowledges the support by the European Research Council (ERC) under the European Union's Horizon 2020 research and innovation programme (Advanced Grant VIRMETAL, grant agreement No. 669141).